\documentclass[sigplan,nonacm]{acmart}

\usepackage{dblfloatfix}

\usepackage{framed}

\input{aux/colors}
\input{aux/code.tex}

\usepackage{cleveref}
\crefname{enumi}{}{}
\Crefname{enumi}{}{}
\crefname{figure}{Figure}{Figures}
\Crefname{figure}{Figure}{Figures}
\crefname{table}{Table}{Tables}
\Crefname{table}{Table}{Tables}
\crefformat{section}{\S#2#1#3}
\crefmultiformat{section}{\S#2#1#3}{ and~#2#1#3}{, #2#1#3}{, and~#2#1#3}

\usepackage{soul}
\newcommand{\TODO}[1]{{\sethlcolor{Mybrightyellow}\hl{TODO: #1}}}

\newcommand{\icc}{\texttt{icc}}

\usepackage{multicol}
\usepackage{multirow}
\usepackage{booktabs}

\renewenvironment{quote}
  {\list{}{\leftmargin=1.5em\rightmargin=1.5em}\item\relax}
  {\endlist}

\begin{document}
\title{AI Coding Agents Need Better Compiler Remarks}

\author{Akash Deo}
\orcid{0009-0000-7585-747X}
\affiliation{%
  \institution{Northwestern University}
  \city{Evanston}
  \state{Illinois}
  \country{USA}
}

\author{Simone Campanoni}
\orcid{0000-0001-9806-7016}
\affiliation{%
  \institution{Northwestern University}
  \city{Evanston}
  \state{Illinois}
  \country{USA}
}

\author{Tommy McMichen}
\orcid{0000-0003-0965-9322}
\affiliation{%
  \institution{Northwestern University}
  \city{Evanston}
  \state{Illinois}
  \country{USA}
}

\thanks{Presented at CoDAIM 2026.}

%%%%%%%%%%%%
% ABSTRACT %
%%%%%%%%%%%%
\begin{abstract}
Modern AI agents optimize programs by refactoring source code to trigger trusted compiler transformations.
This preserves program semantics and reduces source code pollution, making the program easier to maintain and portable across architectures.
However, this collaborative workflow is limited by legacy compiler interfaces, which obscure analysis behind unstructured, lossy optimization remarks that have been  designed for human intuition rather than machine logic.
Using the TSVC benchmark, we evaluate the efficacy of existing optimization feedback.
We find that while precise remarks provide actionable feedback ($3.3\times$ success rate), ambiguous remarks are actively detrimental, triggering semantic-breaking hallucinations.
By replacing ambiguous remarks with precise ones, we show that structured, precise analysis information unlocks the capabilities of small models, proving that the bottleneck is the interface, not the agent.
We conclude that future compilers must expose structured, actionable feedback designed specifically for the future of autonomous performance engineering.

\end{abstract}

\maketitle

%%%%%%%%
% BODY %
%%%%%%%%

\section{Problem Statement} \label{sec:problem}

Large Language Models (LLMs) are rapidly evolving from simple code generators into autonomous performance engineers. 
Recent work demonstrates that these agents can successfully perform complex, source-level refactoring---such as loop splitting and array privatization---to assist compilers in generating highly optimized binaries~\cite{VECTRANS:Zheng:2025,LLM-VECTORIZER:Taneja:2025,CHAIN-OF-VERIFICATION:Kwon:2026,LLM-VERIOPT:Fang:2026}.
However, this collaboration is restricted by a critical bottleneck: the \emph{limited insight} provided by traditional compiler interfaces.

Currently, coding agents operate with an incomplete view of the compiler's analysis. 
While compiler remarks are generally trustworthy, they are often \emph{vague and opaque}, providing little guidance on how to resolve a failed optimization. 
The core issue is not a lack of agent capability, but a lack of \emph{actionable signal}. 
To effectively close the optimization loop, an agent requires feedback that is both \textbf{insightful} (\emph{why} it failed) and \textbf{prescriptive} (\emph{where} and \emph{how} to fix it).

% While precise remarks guide agents to correct solutions, ambiguous remarks actively mislead agents and trigger semantics breaking hallucinations worse than providing no remarks.

\begin{figure}
    \centering
    \includegraphics[width=0.9\linewidth]{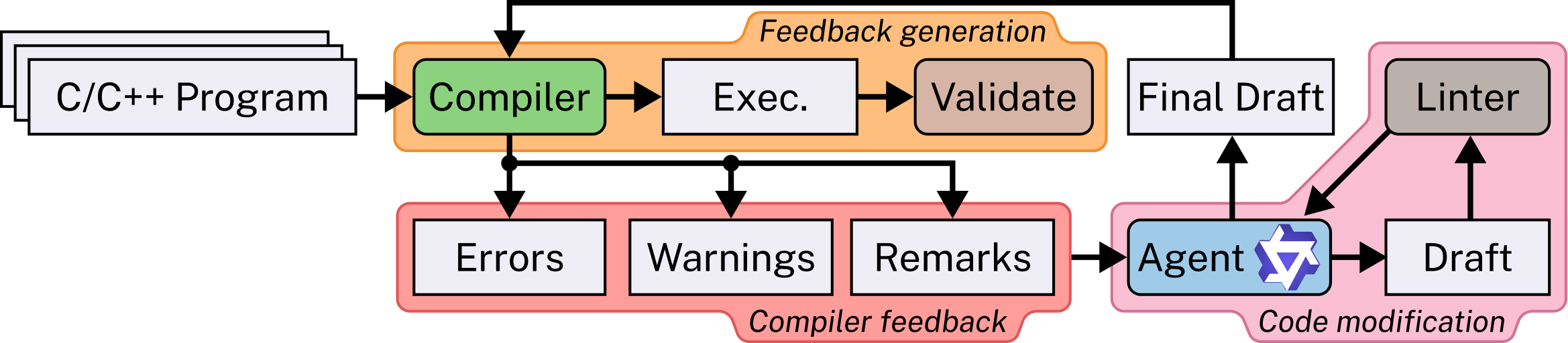}
    \caption{Agentic workflow used in our evaluation.}
    \label{fig:workflow}
\end{figure}

\section{Related Works} \label{sec:related}

\begin{comment}
\textbf{ML for Compiler Optimization.}
Machine learning in compilers has traditionally focused on replacing hand-crafted cost models~\cite{ML-IMPROVE-AUTOVEC:Stock:2012} and predicting optimal vectorization factors~\cite{NEUROVECTORIZER:Haj-Ali:2020}.
Recent research has extended this to utilizing LLMs for generating entire optimization pipelines~\cite{LLM-COMPILER-OPT:Cummins:2023,LLM-COMPILER-FEEDBACK:Grubisic:2024,LLM-COMPILERS:Cummins:2025}.
While effective, these methods largely optimize the compiler's configuration on fixed code.
They overlook the complementary challenge of designing compiler feedback specifically to guide agents in refactoring the source code itself.
\end{comment}

\textbf{LLM-Driven Code Transformation.}
Systems like LLM-Vectorizer~\cite{LLM-VECTORIZER:Taneja:2025} and Astra~\cite{ASTRA-MULTI-AGENT:Wei:2025} bypass compiler safety analyses using low-level intrinsics.
VecTrans~\cite{VECTRANS:Zheng:2025} introduced an iterative loop where an agent is guided by compiler feedback to enable auto-vectorization via source-level transformations.
However, VecTrans treats the compiler's output as a fixed, immutable signal.
This paper shifts the focus to the \emph{quality} of that signal.
We demonstrate that existing feedback is a fundamental bottleneck for AI coding agents and argue for a co-designed interface that exposes deep analytical insights for effective refactoring.

\section{Design and Methodology} \label{sec:system}

We propose an agentic workflow (\cref{fig:workflow}) where an engineer agent is provided with a C/C++ program and comprehensive compiler feedback (warnings, errors, and optimization remarks) and is tasked with performing source-level transformations to enable auto-vectorization.
% \footnotetext{\texttt{clang -fsave-optimization-record} and \texttt{icx -qopt-report=max}.}
We utilize a single-pass evaluation to isolate the impact of diagnostic quality on success rates, rather than measuring trial-and-error in an iterative flow.
The agent refactors based on compiler feedback, with up to three attempts to fix syntax errors. 
We validate by inspecting optimization records for vectorization success and checking semantics via differential testing\footnote{While prior work uses Alive2~\cite{Alive2:Lopes:2021} for formal translation validation, we utilize differential testing as a practical baseline for detecting broken semantics.}.

\section{Evaluation} \label{sec:eval}

% Tommy's earlier draft
% We evaluate Qwen2.5-coder (8B parameters) on TSVC~\cite{TSVC:Siso_Armour_Thiyagalingam:2019}, a benchmark suite consisting of 151 loops designed to test compiler auto-vectorization, running 100 trials per loop for each configuration.
% Critique: We should cite the TSVC-2 GitHub and Maleki et al. 2011 (original TSVC paper) instead of Siso et al. 2019. I failed to mention that we are using TSVC-2, which is the same benchmark for all intents and purposes as TSVC, with functions moved around and some bugfixes here and there. They are very similar. This is supported by the README in the TSVC_2 GitHub repo, and also this paper (RISC-V Vectorization Coverage for HPC: A TSVC-Based Analysis) which says "we conduct a comprehensive instruction coverage analysis ... *using TSVC-2*, an updated version of TSVC with several issues fixed."
% Suggested revision:
We evaluate Qwen2.5-Coder (7B parameters) on TSVC~\cite{TSVC:Maleki_Gao_Gazaran_Wong_Padua:2011}, a benchmark suite consisting of 151 loops designed to test compiler auto-vectorization.
For each loop, we run the workflow under four configurations: \textbf{Clang} 21.1.8 and \textbf{Intel} 2025.3 compilers, both \textbf{with} and \textbf{without} optimization remarks.
We also vary sampling temperature ($T=\{0.2, 0.8, 1.2\}$) to provide insight into the `creativity' needed to enable vectorization, where a higher temperature allows for more stochastic explorations.
We run 100 trials per loop for each configuration.
Our primary goal is to measure how the presence and quality of these remarks influence the agent's success.

% \TODO{Assess alternative beginning for eval section based on dep. remarks. The comment with the alternative starts here.}
\begin{comment}
We evaluate Qwen2.5-Coder (7B parameters) on TSVC~\cite{TSVC:Maleki_Gao_Gazaran_Wong_Padua:2011}, a benchmark suite consisting of 151 loops designed to test compiler auto-vectorization.
For each loop, we run the workflow under five configurations: \textbf{Clang} 21.1.8 and \textbf{Intel} 2025.3 compilers, each \textbf{without remarks}, \textbf{with stock remarks}, and, for Clang, \textbf{with dependence remarks}.
We vary sampling temperature ($T=\{0.2, 0.8, 1.2\}$) to provide insight into the `creativity' needed to enable vectorization. 
We run 100 trials per loop per configuration. Dependence remarks surface Clang's \emph{existing analyses} with a remark detailing dependence type (RAW/WAR/WAW), direction (forward/backward), and source locations, instead of emitting the stock remark "unsafe dependent memory operations." 
For example, we provide "backward write-after-read dependence between source (tsvc.c:1064:18) and destination (tsvc.c:1065:27)."
Our goal is to measure how the presence and quality of remarks influence the agent's success. 
\end{comment}

\textbf{Remarks Improve Success.}
Table 1 shows compiler feedback as the determining factor for success. 
Without remarks, vectorization rates are negligible (<$1.5\%$ for Clang, <$3.7\%$ for Intel) across all temperatures. 
Remarks act as a significant performance multiplier.
At $T=0.8$, providing remarks increases Clang's success rate from $0.80\%$ to $2.68\%$ ($3.3\times$ increase) and Intel's from $2.38\%$ to $6.95\%$ ($2.9\times$ increase).
Notably, at $T=0.2$, the difference in success rate with Intel remarks is much greater than with Clang remarks, suggesting that higher-fidelity diagnostics\footnote{This aligns with the general consensus that the Intel compiler provides more comprehensive and actionable optimization reports than Clang.} reduce the creativity burden on the agent, enabling success at low temperatures.

\textbf{The Benefit of Precision and Cost of Ambiguity.}
For each remark, we aggregate success rates across all benchmarks where that remark appears, shown in \cref{tab:success-by-remark}, which confirms that the utility of feedback is not uniform.
Precise remarks provide massive gains.
For example, identifying an output dependence or anti dependence results in success rate deltas of $+26\%$ and $+15.5\%$, respectively. 
Conversely, vague remarks can be actively \emph{detrimental}.
% Previous: commented out on 3/19
% \TODO{Verify the numbers}
Looking at individual benchmarks, we found that Clang's NonReductionValue remark frequently resulted in semantic hallucinations, where the agent attempted to satisfy a poorly communicated constraint by breaking program logic.
% Clang's NonReductionValueUsedOutsideLoop remark consistently results in negative success rates (up to $-2.42\%$), indicating the agent is more likely to succeed if the feedback is omitted.
% In these cases, the ambiguous signal triggers semantic hallucinations where the agent attempts to satisfy a poorly communicated constraint by breaking program logic.
For example, agents would often incorrectly break loop-carried dependencies when inserting temporaries.
% Akash's change: I'm talking about s1161 here
% Clang's ArrayBounds remark results in negative success rates at high temperatures (up to $-3\%$), indicating creative agents are more likely to succeed if feedback is omitted. 
% In some loops, the ArrayBounds remark is emitted when complex control flow is present. Due to a poorly communicated constraint, the creative agent becomes overly conservative, adding a simple compiler directive (e.g., \texttt{\#pragma clang loop vectorize(enable)}) instead of restructuring control flow. 

\textbf{Temperature Sensitivity.}
Success rates generally scale with temperature, confirming that satisfying complex vectorization constraints often requires the stochastic `creativity' found at $T=1.2$, whereas those same solutions are pruned at low temperatures.
% Tommy's draft
% However, certain structural fixes, such as resolving \texttt{multiple\_exits}, peak at $T=0.2$ ($+10.61\%$) and degrade as temperature increases.
% Revised draft
However, certain structural fixes, such as resolving Intel's MultipleExits remark, peak at $T=0.2$ ($+22.33\%$) and degrade as temperature increases.
This suggests that, with compiler feedback, structural fixes can be safely performed using deterministic reasoning.
% while semantic refactoring (e.g., scalar expansion) requires higher temperature, structural refactoring is best handled by deterministic reasoning.

\textbf{Precise Remarks.}
% \TODO{Replace this data with new, better remarks data. }
% \TODO{State that this is information that the compiler needs to compute anyways to know it can/can't apply the optimization when it emits the remark}
To investigate how existing remarks can be improved, we hand-write precise remarks, providing detailed insights for why vectorization failed.
In particular, we expose the underlying dependence analysis instead of producing a detrimental ``unsafe dependent memory operations in loop'' remark in Clang.
% Tommy's draft
% For instance, we provide "assumed write-after-read dependence between a[i+1] [tsvc.c:1148:13] and a[i] [tsvc.c:1149:27]."
% Revised draft
For instance, we provide:
\begin{quote}
\textit{assumed write-after-read dependence between \\
        a[i+1] [tsvc.c:1148:13] and a[i] [tsvc.c:1149:27] \\
        suggestion: consider using a temporary to store a[i+1]}
\end{quote}
With precise remarks, success rates drastically improve: identifying write-after-read dependencies achieves $+45\%$ at $T=0.2$, while read-after-write shows $+50\%$ at $T=0.8$.
These gains demonstrate that exposing precise dependence information enables agents to reliably apply transformations.
% Commented out on 3/19
% Write-after-write shows inconsistent results across temperatures, suggesting that dependence information alone is insufficient.
Write-after-write, by contrast, shows only modest gains, suggesting that dependence information alone is insufficient. 
Additional analysis may be needed to make this remark actionable.

% \TODO{Assess whether this can replace precise remarks. Comment starts here}
\begin{comment}
\textbf{Dependence Remarks.}
With dependence remarks, success rates drastically improve: identifying Backward WAW ($+31.00\%$ at $T=0.2$) and Backward WAR ($+10.83\%$ at $T=1.2$) are the largest deltas of any remark.
These two categories also exhibit opposite temperature profiles: WAW peaks at low temperature while WAR peaks at high temperature, suggesting the nature of the dependence determines whether the fix requires deterministic reasoning or creative exploration.
Even 'Unknown' remarks (which provides partial dependence information) achieve $+6.67\%$ at $T=1.2$.
Backward RAW remarks are harmful across all temperatures, suggesting that additional analysis may be needed to make the remark actionable. 
\end{comment}

\begin{table}
\centering

\begin{tabular}{l l | c c c}
\toprule
\textbf{Compiler} & \textbf{Config} & \textbf{T=0.2} & \textbf{T=0.8} & \textbf{T=1.2} \\
\midrule
Clang & No Remarks & $0.20\%$ & $0.80\%$ & $1.45\%$ \\
Clang & Remarks    & $0.64\%$ & $2.68\%$ & $3.93\%$ \\
\midrule
Intel & No Remarks & $1.10\%$ & $2.38\%$ & $3.67\%$ \\
Intel & Remarks    & $4.59\%$ & $6.95\%$ & $7.83\%$ \\
\bottomrule
\end{tabular}

\caption{Success rate of vectorization using different compilers with varying temperatures over 100 trials. Benchmarks that can be vectorized without refactoring are excluded.}
\label{tab:results}
\end{table}

\begin{table}
\centering
\begin{tabular}{l l | r r r}
\toprule
\multicolumn{2}{c}{\textbf{Remark}} & \textbf{T=0.2} & \textbf{T=0.8} & \textbf{T=1.2} \\
\midrule
\multirow{7}{*}{\rotatebox{90}{\textbf{Intel}}}
&Output Dependence & +11.00 & \textbf{+26.00} & +10.00 \\
&Anti Dependence & +5.50 & \textbf{+15.50} & +5.50 \\
&Multiple Exits & \textbf{+22.33} & +11.67 & +11.33 \\
&Flow Dependence & +2.38 & \textbf{+4.38} & +4.27 \\
&Function Call & 0.00 & 0.00 & +0.33 \\
&Loop Control Var & 0.00 & +1.00 & +1.67 \\
% &Outer Loop & 0.20 & 2.00 & 2.70 \\
% &\TODO{Remove?}Seems Inefficient & 0.00 & 0.00 & 0.00 \\
\midrule
\multirow{6}{*}{\rotatebox{90}{\textbf{Clang}}}
&ArrayBounds & \textbf{+7.00} & -2.00 & -3.00 \\
&EarlyExit & 0.00 & \textbf{+12.00} & +7.00 \\
&NotBeneficial & 0.00 & +4.50 & \textbf{+7.50} \\
&UnsafeDependency & -0.08 & +4.50 & \textbf{+6.25} \\
&Libcall/Instr & 0.00 & -2.50 & -1.00 \\
% &MissedDetails & 0.43 & 1.89 & 2.48 \\
&NonReductionValue & 0.00 & +1.00 & +1.23 \\
% &NotPossible & 0.80 & 2.92 & 3.28 \\
% &UncountableLoop & 5.00 & 9.00 & 5.00 \\
\midrule
% Commented out on 3-19
\multirow{3}{*}{\rotatebox{90}{\textbf{Precise}}}
& ReadAfterWrite & 0.00 & \textbf{+50.00} & +59.00 \\
& WriteAfterRead & \textbf{+45.00} & +40.80 & +35.20 \\
& WriteAfterWrite & 0.00 & +9.00 & +7.00 \\
% \midrule
% \multirow{4}{*}{\rotatebox{90}{\textbf{Dep.}}}
% &BackwardVectorizable RAW & 0.00 & 0.00 & 4.00 \\
% &Backward RAW & -1.00 & -0.50 & -1.00 \\
% &Backward WAR & 5.17 & 10.33 & \textbf{10.83} \\
% &Backward WAW & \textbf{31.00} & 9.00 & 8.00 \\
% &CantIdentifyArrayBounds & 18.50 & 3.50 & 0.00 \\
% &CantVectorizeInstruction & 0.00 & -2.50 & -0.50 \\
% &CantVectorizeLibcall & 0.00 & -2.50 & -0.50 \\
% &Forward RAW & 0.00 & 0.00 & -1.00 \\
% &Forward WAR & 7.75 & 5.50 & 7.75 \\
% &Indirect Unsafe & 0.00 & 0.00 & 2.00 \\
% &MissedDetails & 1.52 & 2.52 & 3.05 \\
% &NonReductionValueUsedOutsideLoop & 0.00 & 1.42 & 1.90 \\
% &NotBeneficial & 0.00 & 6.00 & 6.50 \\
% &NotPossible & 1.52 & 4.12 & 4.52 \\
% &Unknown hazard & 0.33 & 6.33 & \textbf{6.67} \\
% &UnsafeDep & 2.50 & 6.67 & 7.08 \\
% &UnsupportedUncountableLoop & 0.00 & 1.00 & 2.00 \\
% &WritesInEarlyExitLoop & 0.00 & 7.00 & 15.00 \\
\bottomrule
\end{tabular}

\caption{Difference in success rate with and without remarks, aggregated by remark type.}

\label{tab:success-by-remark}
\end{table}

\section{Discussion and Conclusion} \label{sec:discussion}

Agents thrive on precise, data-flow-level signals, such as explicit dependencies with source-level debug information, while collapsing under ambiguous structural warnings.
Exposing \emph{why} the transformation failed allows the agent to distinguish between genuine data hazards and conservative static approximations.

We demonstrate that the bottleneck for AI-driven optimization is not model capability, but the opacity of traditional compiler interfaces. 
Open questions in this area include addressing thinking models: future work must investigate whether compiler static analysis can serve as a lower-cost, higher-accuracy replacement for test-time scaling techniques like agent thinking and whether thinking models might amplify creativity benefits from better optimization remarks.
% Precise diagnostics act as a performance multiplier, whereas vague signals provoke hallucinations that break semantics.
By evolving from purely descriptive observations to structured, prescriptive analysis, we transform the AI agent from a stochastic code generator into a reliable performance engineer capable of navigating complex, semantics-preserving optimizations.

\bibliographystyle{ACM-Reference-Format}
\bibliography{bib/references}

\end{document}